# Data Mining for Actionable Knowledge: A Survey[*]

Zengyou He, Xiaofei Xu, Shengchun Deng

Department of Computer Science and Engineering,Harbin Institute of Technology ,

92 West Dazhi Street, P.O Box 315, Harbin 150001, P. R. China

zengyouhe@yahoo.com, {xiaofei, dsc}@hit.edu.cn

**Abstract** The data mining process consists of a series of steps ranging from data cleaning, data selection and transformation, to pattern evaluation and visualization. One of the central problems in data mining is to make the mined patterns or knowledge actionable. Here, the term actionable refers to the mined patterns suggest concrete and profitable actions to the decision-maker. That is, the user can *do* something to bring direct benefits (increase in profits, reduction in cost, improvement in efficiency, etc.) to the organization's advantage. However, there has been written no comprehensive survey available on this topic. The goal of this paper is to fill the void.

In this paper, we first present two frameworks for mining actionable knowledge that are inexplicitly adopted by existing research methods. Then we try to situate some of the research on this topic from two different viewpoints: 1) data mining tasks and 2) adopted framework. Finally, we specify issues that are either not addressed or insufficiently studied yet and conclude the paper.

## 1. Introduction

Data mining, as well as its synonyms knowledge discovery and information extraction, is frequently referred to the literature as the process of extracting interesting information or patterns from large databases. There are two major issues in data mining research and applications: patterns and interest. The techniques of pattern discovering include classification, association, outlier and clustering. Interest refers to patterns in business applications being useful or meaningful. Other than general measures and domain specific measures, an important measure of the interestingness is whether it can be used in the decision making process of a business to increase its profit. Data mining may also be viewed as the process of turning the data into information, the information into action, and the action into value or profit [15, 16]. That is, mining those *actionable* patterns that the user can act on them to his advantage.

Extensive research in data mining has been done on discovering patterns about the underlying data. Despite such phenomenal success, most of these techniques stop short of the final objective of data mining-providing possible actions to maximize profit while reducing costs. While these techniques are essential to move the data mining result to the eventual application, they nevertheless require a great deal of expert manual to post-process mined pattern. Most of the post-processing techniques have been limited to producing visualization results, but they do not directly suggest actions that would lead to the increase on the objective function such as profit.

[*] This work was supported by the High Technology Research and Development Program of China (No. 2003AA4Z2170, No. 2003AA413021) and the IBM SUR Research Fund.

Overall, taking actions to make profit for individual or organization is the ultimate goal of data mining. Thus, some recent research [e.g. 1, 3-38] has focused more on the actionable knowledge data mining problem. However,there has been no comprehensive review of  actionable knowledge data mining.

In this paper, we attempt to fill the void by presenting a summary of research on this topic. Firstly, we present two frameworks for mining actionable knowledge that are inexplicitly adopted by existing research methods, namely tightly coupled framework and loosely coupled framework.

Secondly, we try to situate some of the research on this topic from two different viewpoints: 1) data mining tasks and 2) adopted frameworks. Finally, we specify issues that are either not addressed or insufficiently studied yet and conclude the paper.

This paper is organized as follows: Section 2 discusses the concept of actionable knowledge. Section 3 presents two high-level data mining frameworks for actionable knowledge acquisition. Then a summary of research on this topic from two different viewpoints will be reviewed in Sections 4.Finally, Section 5 discusses future research directions.

## 2.  The Concept of Actionability

According to actionability measure, a pattern is interesting because the user can do something about it; that is, the user can react to it to his or her advantage [1]. Actionability is an important subjective measure of interestingness because users are mostly interested in the knowledge that permits them to do their jobs better by taking some specific actions in response to the newly discovered knowledge.

As pointed out in [1], actionability is really the key concept that is of main interest to business people. However, it is an elusive concept that is very difficult to capture formally for the following reasons [1]. First, we have to partition the space of all the patterns into a finite (and hopefully small) set of equivalence classes and associate an action or a class of actions with each equivalence class. Since in many cases we do not know the space of all the patterns, this task can simply be infeasible. Furthermore, even if we know this space, the partitioning it into equivalence classes and assignments of actions for each class can be a very complicated task. Secondly, even if we succeed in associating or actions with classes of patterns, actions and the mapping of actions to patterns may often change over time, and the task of reassigning patterns to actions can be a very hard one.

Despite such phenomenal difficulties, most of previous techniques have successfully identified actionable knowledge either by integrating the mining process into decision process or utilizing special business objective, as will be reviewed in later sections.

## 3.  Two Frameworks

Usually, two frameworks for mining actionable knowledge are inexplicitly adopted by existing research methods: 1) loosely coupled framework and 2) tightly coupled framework.

In data mining process, discovering patterns from the underlying data is usually not a goal in itself, but rather a means to an end. The actual goal is to determine concrete and profitable actions to the decision-maker. Usually, a data mining algorithm is executed first and then, on the basis of these data mining results, the profitable actions are determined. Hence, in the loosely coupled framework, extraction of actionable knowledge is preceded by some particular data mining task, i.e., they are two loosely coupled process, as shown in Fig.1.

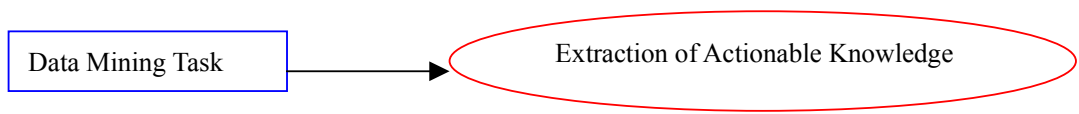

**Fig.1** The general procedure to go from data mining task to actionable knowledge in loosely coupled framework

Apparently, the main advantage of loosely coupled framework is its flexibility and independencies on applications. However, no guarantee that the discovered patterns in the first step will lead to actionable knowledge that are capable of maximizing profits.

In tightly coupled framework, decision-making task is seamless integrated into the data mining task, therefore leads to the formulation of a new data mining or optimization problem, as shown in Fig.2. In contrary to loosely coupled framework, two tasks (data mining and decision making) are considered in one unifying framework, which determines the optimal mined patterns and the optimal actions using the same criterion. Hence, tightly coupled framework is better than loosely coupled framework in finding actionable knowledge to maximize profits. However, it deserves the following disadvantages: 1) It is strongly dependent on the application domain, 2) the new formulated problem is usually very complex and 3) defining and solving the new data mining problem is also a non-trivial task.

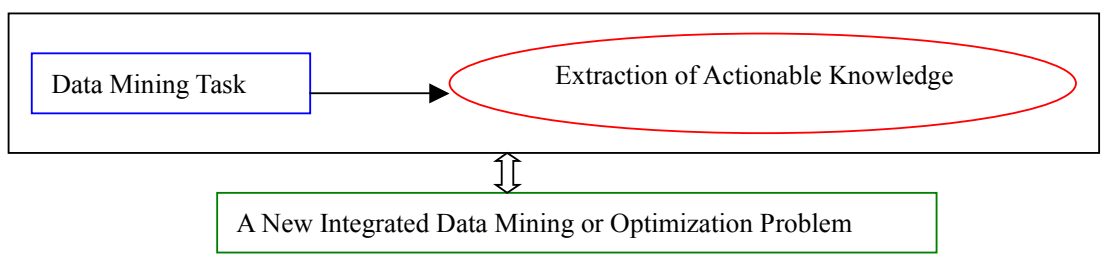

**Fig.2** The general procedure to go from data mining task to actionable knowledge in tightly coupled framework – A seamless integration methodology

## 4. Surveys from Two Different Viewpoint

In this section, we try to situate some of the research on this topic from two different viewpoints: 1) data mining tasks and 2) adopted frameworks.

### 4.1 The Viewpoint of Data Ming Tasks and Algorithms

Data mining tasks and algorithms refer to the essential procedure where intelligent methods are applied to extract useful patterns. There are many data mining tasks such as clustering, classification, associations, outlier detection etc. Each task can be thought as a particular kind of problem to be solved by a data mining algorithm. Generally there are many different algorithms could serve the purpose of the same task. Meanwhile, some algorithms can be applied to different tasks. In this section, we review current researches on mining actionable knowledge from the viewpoint of data mining tasks and algorithms. We basically organize the review according to different tasks. The tasks we discussed are association rules, clustering, classification, and outlier detection.

**4.1.1 Association Rules**

The task of association rule mining is to find certain association relationships among a set of objects (called items) in a database. The association relationships are described in association rules. Each rule has two measurements, support and confidence. Confidence is a measure of the rule's strength, while support corresponds to statistical significance.

The task of discovering association rules was first introduced in 1993 [2]. Originally, association rule mining focused on market "basket data" which stores items purchased on a per-transaction basis. A typical example of an association rule on market "basket data" is that 70% of customers who purchase bread also purchase butter. Later, association rule mining was also extended to handle quantitative data.

In the literature of data mining, there are a lot of studies on designing scalable algorithms for mining association rules. Such studies are particularly useful with a large amount of data in order he to understand the customer behavior in their stores. However, it is generally is true that the results of association rule mining are not directly useful for the business sector [6]. Therefore there has been research in examining more closely the business requirements and finding solutions that are suitable for particular issues, such as marketing and inventory control. Recently, some researchers begin to study the utility of association on aiding decision making for revenue-maximizing enterprises [3-12].

The problem of optimal product selection with association rules is considered in [3-6]. The problem is that in a typical retail store, the types of products should be refreshed regularly so that losing products are discarded and new products are introduced. Hence, it is interested to find a subset of the products to be discontinued so that the profit can be maximized. The formulation of the problem considers the important factor of cross selling which is the influence of some products on the sales of other products. The cross-selling factor is embedded into the decision calculation of the maximum profit gain from a decision. In those models, association rules are used to model the cross-selling effect among items.

In [7], association rules are used to generate recommendation actions. In their model, given a set of past transactions and pre-selected target items, and it intends to build a model for recommending target items and promotion strategies to new customers, with the goal of maximizing the net profit.

The researches [8,9] address ways to generate *interesting* patterns by incorporating managers' prior knowledge in the process of searching for patterns in data. Specifically, they focus on providing methods that generate *unexpected* patterns with respect to managerial intuition by eliciting managers' beliefs about the domain and using these beliefs to seed the search for unexpected patterns in data.

Association rules are mined from transaction databases with the goal of improving sales and services. Two standard measures called support and confidence are used for mining association rules. However, both measures are not directly linked to the use of association rules in the context of marketing. In order to resolve this problem, Lin et al. [10] consider a framework for value added association rules by attaching numerical values to itemsets, representing profits, importance, or benefits of itemsets.

Liu et al. [11] consider the problem of post-processing association rules to prune non actionable associations, so that the final size of association rules are greatly reduced.

In [12], the authors present a new approach, called Objective-Oriented utility-based Association (OOA) mining, to modeling such association patterns that are explicitly relating to a user's objective and its utility. Due to its focus on a user's objective and the use of objective utility as key semantic information to measure the usefulness of association patterns, decision-making task is seamless integrated into the association mining process.

**4.1.2 Classification**

Classification problems aim to identify the characteristics that indicate the group to which each instance belongs. Classification can be used both to understand the existing data and to predict how new instances will behave. Classification is a well-recognized data mining operation that has been studied extensively in the fields of statistics, pattern recognition, decision theory, machine learning literature, neural network, etc.

Recently, it is argued that classification in real business applications (especially in marketing and customer relationship management) is not enough, some further actions should to taken to re-classify some customers from less desired decision class to the more desired one.

To that end, the notion of an action rule was proposed in [13] and further extended in [14]. It is assumed that attributes in a database are divided into two groups: stable and flexible. By stable attributes we mean attributes whose values cannot be changed (age, marital status, number of children are the examples). On the other hand, attributes (like percentage rate or loan approval to buy a house in certain area) whose values can be changed or influenced are called flexible. Rules are extracted from a decision table given preference to flexible attributes. In general, any action rule provides hints to a business user what changes within flexible attributes are needed to re-classify some customers from a lower profitability class to a higher one.

Ling et al [15] and Yang et al [16] discussed the techniques on post-processing decision tress to maximizing expected net profit. They also consider how to take some actions to re-classify some customers from less desired decision class to the more desired one. More importantly, they explicitly incorporating the cost of each action in maximizing expected net profit.

Although decision tree is employed in [15,16] and a rough set based classification is utilized in [13,14], despite their apparent difference in choosing classification technique, both of them can be regarded as classification-based action rule mining technique, i.e., post-processing classification results to get actionable knowledge.

Yang et al [17] presented an approach to use 'role models' for generating advice and plans. These role models are typical cases that form a case base and can be used for customer advice generation. For each new customer seeking advice, a nearest-neighbor algorithm is used to find a cost-effective and highly probable plan for switching a customer to the most desirable role models.

Elovici and Braha [18] develop a decision-theoretic framework for evaluating data mining systems, which employ classification methods, in terms of their utility in decision-making. The decision-theoretic model provides an economic perspective on the value of "extracted knowledge, in terms of its payoff to the organization, and suggests a wide range of decision problems that arise from this point of view. The relation between the *quality* of a data mining system and the amount of investment that the decision maker is willing to make is formalized.

Class association rule (CAR) is a small subset of association rules whose right hand sides are restricted to the class label. *CAR* was first proposed by Bing Liu, et al. in 1998 [19]. They integrate the techniques of association rule mining and classification rule mining by focusing on mining a special subset of association rules. Then the discovered *CARs* are used to build a classifier. Li et al [20] proposed another algorithm by using multiple class association rules. Actually, the purpose of these two approaches is to construct classifiers, rather than finding interesting re-classification rules.

**4.1.3 Clustering**

The problem of clustering data arises in many disciplines and has a wide range of applications. Intuitively, the clustering problem can be described as follows: Let *W* is a set of n data points in a multidimensional space. Find a partition of *W* into classes such that the points within each class are *similar* to each other, while points in different class are *dissimilar* to each other.

One of the main motivations of clustering has been the hope that, by clustering the data in meaningfully distinct clusters, we can then proceed to make independent decisions for each cluster [21]. To our knowledge, Cleinberg et al. [21] provide the first formalism of clustering that explicitly embodies this motivation. The framework formulated in [21] suggests a number of interesting computational issues, related to sensitivity analysis and clustering. While they integrated the decision problem into the clustering process, hence their formulation is limited to a particular problem and is not applicable to more general decision problems.

Abidi et al. [22, 23] presented a symbolic rule extraction workbench that leverages rough sets theory to inductively extract CNF form symbolic rules from data clusters. Aggarwal et al. [24 ] discussed a technique for discovering localized association in segments of data using clustering, and aimed to expose a customer pattern that is more specific than the aggregated behavior. The research in both [22,23] and [24] mainly focused on extracting rules form data clusters after a clustering procedure, in hope that profitable actions can be made to each cluster with the help of those cluster-defining rules.

Contrast-set mining is a new data mining technique designed specifically to identify differences between contrasting groups from observational multivariate data [25]. The problem of mining contrast sets is formulated as: finding conjunctions of attributes and values that differ especially meaningfully in their distribution across groups, that is, the conjunctions of attributes and values are *significant* and *large*. Obviously, applying contrast sets mining to data clusters will enable us to understand the differences between clusters and make independent decisions for each cluster. Motivated by the above observation, He et al. [26] use the contrast set mining technique to mine cluster-defining actionable rules from data clusters. In that way, it is easy to understand the differences between clusters and provide insights into the complex inter-relationships between the various data clusters and directly suggests actions for decision makers.

Simultaneously, Zhang et al. [27] study market share rules, rules that have a certain market share statistic associated with them. Such rules are particularly relevant for decision making from a business perspective. Motivated by market share rules, they also consider statistical quantitative rules (SQ rules) that are quantitative rules in which the RHS can be any statistic that is computed for the segment satisfying the LHS of the rule. In essence, both [26] and [27] aim at mining cluster-defining rules that have statistical significance to understand the differences between clusters, so as to directly suggests actions for decision makers.

In some commercial software, such as SPSS, a decision tree algorithm is utilized to produce cluster rules after cluster analysis for making decisions. Similarly, decision tree algorithm is used for producing cluster rules after segmentation of stock trading customers in [28].

In [29], a new formulation of the conceptual clustering problem is proposed where the goal is to explicitly output a collection of simple and meaningful conjunctions of attributes that define the clusters. With conjunctions of attributes as the description of clusters, decision maker can take actions.

The authors in [30] develop a novel market segmentation methodology based on product specific variables such as purchased items and the associative monetary expenses from the transactional history of customers. After completing segmentation, a designated RFM model is used to analyze the relative profitability of each customer cluster to find cluster-defining actionable knowledge.

In [31], a joint optimization approach is presented to address the segmentation of customers and determining the optimal policy (i.e. what action to take from a set of available actions) towards each segment in a unifying framework. Hence, it shares some common ideas with [21]. However, the formulation in [31] is more meaningful in practice since it determines which marketing policy is optimal using markov decision processes.

**4.1.4 Outlier Detection**

In contrast to traditional data mining task that aims to find the general pattern applicable to the majority of data, outlier detection targets the finding of the rare data whose behavior is very exceptional when compared with rest large amount of data. Studying the extraordinary behavior of outliers helps uncovering the valuable knowledge hidden behind them and aiding the decision makers to make profit or improve the service quality. Thus, mining for outliers is an important data mining research with numerous applications, including credit card fraud detection, discovery of criminal activities in electronic commerce, weather prediction, and marketing.

A well-quoted definition of outliers is firstly given by Hawkins [32]. This definition states, "An outlier is an observation that deviates so much from other observations as to arouse suspicion that it was generated by a different mechanism". With increasing awareness on outlier detection in data mining literature, more concrete meanings of outliers are defined for solving problems in specific domains. Nonetheless, most of these definitions follow the spirit of the Hawkins-Outlier.

However, most existing studies on outliers focus only on the identification of outliers, how to find useful actionable knowledge from the results of outlier mining is not fully addressed, with only a few exceptions [33-38].

Knorr and Ng [33-34] consider providing intensional knowledge of the outliers--- a description or an explanation of why an identified outlier is exceptional. They propose finding strongest and weak outliers and their corresponding structural intensional knowledge. As pointed in [33-34], intensional knowledge helps the user to: (i) evaluate the validity of the identified outliers, and (ii) improve one's understanding of the data.

In [35], Chen et al. study in a general setting the notion of outlying patterns as intentional knowledge of outliers. They propose a model for defining outlying patterns with the help of categorical and behavioral similarities of outliers, and efficient algorithms for mining knowledge sets of distance based outliers and outlying patterns.

On the other hand, the problem of class outlier detection is considered in [36-38]. That is, given a set of observations with class labels, find those that arouse suspicions, taking into account the class labels.

In [36], the class outlier is defined as "semantic outlier". A semantic outlier is a data point, which behaves differently with other data points in the same class, while looks 'normal' with respect to data points in another class. In contrast, the authors of Ref. [37] propose the cross-outlier detection problem. That is, given two sets (or classes) of objects, find those that deviate with respect to the other set.

In [38], the notion of class outlier is proposed by generalizing works in [36,37]. How to apply class outliers in decision-making process of analytical CRM (customer relationship management) is discussed and its effectiveness is evaluated on real datasets.

## 4.2 The Viewpoint of Adopted Framework

As discussed in Section 3, two frameworks for mining actionable knowledge are usually inexplicitly adopted by existing research methods: loosely coupled framework and tightly coupled framework. Table 1 depicts our classification on existing researches with respect to different frameworks and different data mining tasks.

**Table 1** Classification on existing researches with respect to different frameworks and different data mining tasks

| Task \ Framework | Association | Clustering | Classification | Outlier |
|---|---|---|---|---|
| loosely coupled framework | [10], [11] | [22-28], [30] | [13-17] | [33-35] |
| tightly coupled framework | [3-9], [12] | [21],[29],[31] | [18] | [36-38] |

From Table 1, some important observations are summarized as follows.

(1) For those description-oriented data mining tasks, such as association and clustering, tightly coupled framework is more favorable and widely adopted, because moving the results of description-oriented data mining task to the eventual application in the loosely coupled framework is very difficult. On the contrary, those prediction-oriented data mining tasks, such as classification, are more likely to adopt loosely coupled framework due to their mining results are more easier to be deployed in application.

(2) Recently, we have witnessed that data mining and operations research (and optimization techniques) are beginning to intersect and interact more frequently. While data mining can assist in the automated discovery of actionable insights from data, the efficient execution of the actions can only be effected by coupling the output of data mining with optimization methods [39]. That is, only applying optimization techniques to the outputs of data mining can effectively make decisions. Hence, it is not strange that most researches in tightly coupled framework have formulated the new problem as an optimization problem.

(3) In loosely coupled framework, the techniques used for post-processing previously mined patterns varied significantly. This is because this framework is less dependent on applications. Hence, to generate more general actionable knowledge, all techniques have their own rights.

## 5. Discussions and Conclusions

This paper presents two frameworks for mining actionable knowledge that are inexplicitly adopted by existing research methods. Consequently, we situate some of the researches on this topic from two different viewpoints: 1) data mining tasks and 2) adopted framework.

However, making the mined patterns or knowledge actionable is also not the ultimate goal of data mining. The final objective is to deploy actions to maximize profit while reducing costs. As we have noted, coupling the output of data mining with optimization methods is the most feasible way to determine optimal actions to increase profit. Hence, further research on mining actionable knowledge should pay more attention on the interaction between data mining and optimization technique. Two recent surveys [40,41] on the intersection and interaction between these two fields are already available.